\documentclass{article}

\usepackage{arxiv}

\usepackage[utf8]{inputenc} 
\usepackage[T1]{fontenc}    
\usepackage{hyperref}       
\usepackage{url}            
\usepackage{booktabs}       
\usepackage{amsfonts}       
\usepackage{nicefrac}       
\usepackage{microtype}      
\usepackage{lipsum}		
\usepackage{graphicx}
\usepackage{natbib}
\usepackage{wrapfig}
\usepackage{doi}
\usepackage{amsmath} 

\newcommand{\angstrom}{\textup{\AA}}
\renewcommand{\phantom}{} 

\title{SHREC 2021:\\Classification in cryo-electron tomograms}

\author{\parbox{\textwidth}{\centering
\thanks{For any questions, please contact by e-email: \texttt{i.gubins@uu.nl}}
\thanks{Conference version of the paper can be found here: \url{https://diglib.eg.org/handle/10.2312/3dor20211307}}
Ilja Gubins$^{1,2}$,
Marten L. Chaillet\footnotemark[1]$^{2}$,
Gijs van der Schot\footnotemark[1]$^{2}$,
M. Cristina Trueba$^{2}$,
Remco C. Veltkamp\footnotemark[1]$^{1}$,
Friedrich F\"orster\footnotemark[1]$^{2}$,
Xiao Wang$^{3}$,
Daisuke Kihara$^{4,3}$,
Emmanuel Moebel$^{5}$,
Nguyen P. Nguyen$^{6}$,
Tommi White$^{7}$,
Filiz Bunyak$^{6}$,
Giorgos Papoulias$^{8}$,
Stavros Gerolymatos$^{8}$,
Evangelia I. Zacharaki$^{8}$,
Konstantinos Moustakas$^{8}$,
Xiangrui Zeng$^{9}$,
Sinuo Liu$^{9}$,
Min Xu$^{9}$,
Yaoyu Wang$^{10}$,
Cheng Chen$^{10}$,
Xuefeng Cui$^{{10}}$,
Fa Zhang$^{{11}}$
}
\\
{\parbox{\textwidth}{\vspace{2em}
$^1$ Department of Information and Computing Sciences, Utrecht University, Netherlands\\
$^2$ Department of Chemistry, Utrecht University, Netherlands\\
$^3$ Department of Computer Science, Purdue University, USA\\
$^4$ Department of Biological Sciences, Purdue University, USA\\
$^5$ Inria Rennes Bretagne Atlantique, France\\
$^6$ Department of Electrical Engineering and Computer Science, University of Missouri, USA\\
$^7$ Department of Biochemistry \& Electron Microscopy Core Facility, University of Missouri, USA\\
$^8$ Department of Electrical and Computer Engineering, University of Patras, Greece\\
$^9$ Computational Biology Department, Carnegie Mellon University, USA\\
$^{10}$ School of Computer Science and Technology, Shandong University, China\\
$^{11}$ Institute of Computing Technology, Chinese Academy of Sciences, China
}
}
}

\date{}

\renewcommand{\headeright}{}
\renewcommand{\undertitle}{}
\renewcommand{\shorttitle}{SHREC 2021: CryoET Track}

\hypersetup{
pdftitle={SHREC 2021: CryoET Track},
pdfsubject={cs.CV},
pdfauthor={I. Gubins, Marten L. Chaillet et al.},
pdfkeywords={Cryo-electron tomography, Benchmark, Localization, Classification},
}

\begin{document}
\maketitle

\begin{abstract}
Cryo-electron tomography (cryo-ET) is an imaging technique that allows three-dimensional visualization of macro-molecular assemblies under near-native conditions. Cryo-ET comes with a number of challenges, mainly low signal-to-noise and inability to obtain images from all angles. Computational methods are key to analyze cryo-electron tomograms.

To promote innovation in computational methods, we generate a novel simulated dataset to benchmark different methods of localization and classification of biological macromolecules in tomograms. Our publicly available dataset contains ten tomographic reconstructions of simulated cell-like volumes. Each volume contains twelve different types of complexes, varying in size, function and structure.

In this paper, we have evaluated seven different methods of finding and classifying proteins. Seven research groups present results obtained with learning-based methods and trained on the simulated dataset, as well as a baseline template matching (TM), a traditional method widely used in cryo-ET research. We show that learning-based approaches can achieve notably better localization and classification performance than TM. We also experimentally confirm that there is a negative relationship between particle size and performance for all methods.
\end{abstract}


\section{Introduction}
\label{sec:introduction}

Cryo-electron tomography (cryo-ET) is an application of transmission electron microscopy, in which biological samples are cryogenically vitrified and imaged as they are sequentially tilted. The resulting ``tilt-series'' of 2D projections can be merged into a 3D reconstruction. Cryo-electron tomograms feature macromolecular assemblies in their cellular context, offering insight into life processes at its smallest scale~\cite{yahav2011cryo}. This data is key for improving our understanding and determining modes of actions of drugs.

\begin{wrapfigure}{l}{0.5\textwidth}
    \centering
    \includegraphics[width=.5\textwidth]{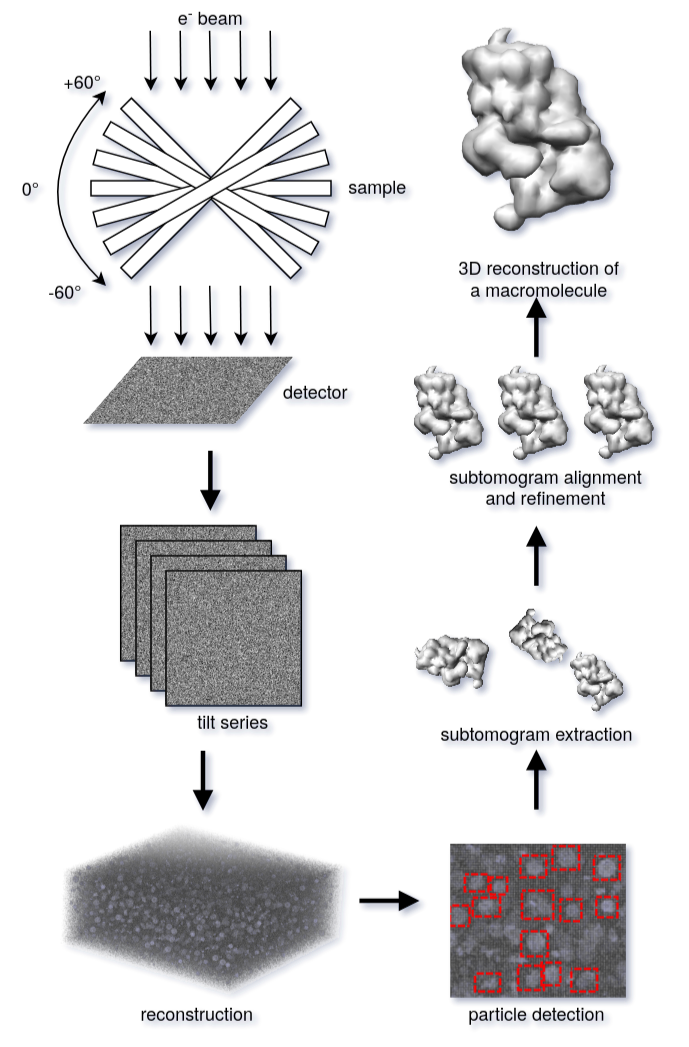}
    \caption{The overall process of cryo-electron tomography from data collection to reconstruction and subtomogram averaging.}
    \label{fig:cryoet_process}
\end{wrapfigure}

The approach comes with a number of challenges. Imaging electrons strongly interact with biological matter, severely limiting the possible dose to avoid damaging the sample during imaging. The limited dose in turn limits signal-to-noise and effective resolution of the tomograms to about $5 nm$ ($50 Å$). Such resolution is not enough to distinguish structural details of biomolecules. The common approach to increase resolution of the biomolecule of interest is subtomogram averaging~\cite{pfeffer2018unravelling}. It involves aligning and averaging copies of the same particles, introducing the challenge of correctly localizing and identifying those particles in the raw tomogram~(Figure~\ref{fig:cryoet_process}).

Another major challenge is the constraint on imaging angles, up to $\pm60^\circ$ due to sample thickness, resulting in an incomplete reconstruction with a ``missing wedge'' in Fourier space. Manual analysis of such data is rarely feasible and often provides subjective results, leading to the interest in automated approaches. The most common of such approaches, for biological particles of known structures, is template matching \cite{frangakis2002identification}. Cross correlation between the template and the entire tomogram indicates locations and angles, where the template fits the best. For particles with unknown structures, reference-free methods must be used. The most common approach is based on applying Difference of Gaussian (DoG) \cite{voss2009dog}: a band-pass filter that removes noisy high frequency components and homogeneous low frequency areas, obtaining borders of the structures.

In recent years, machine learning has seen successful application to cryo-ET. Classical support vector machines have been used for both detection and classification \cite{chen2012detection}. With ever increasing amounts of data captured by cryo-EM and -ET methods \cite{baldwin2018big}, deep learning methods are gaining popularity. Supervised methods were proposed for localization \cite{wang2016deeppicker}, classification \cite{che2018improved}, end-to-end segmentation \cite{chen2017convolutional} and joint localization and classification \cite{li2019automatic}, providing faster and often more accurate results than template matching \cite{gubins2019classification}. Moreover, methods based on clustering of representational features \cite{xu2019novo}, segmentation by manually designed rules \cite{xu2013automated} and geometric matching \cite{zeng2020gum} provide unsupervised and weakly-supervised alternatives, reducing the dependency on annotated data.

Each of the mentioned methods is validated on different tasks and different datasets (i.e., data acquisition parameters and microscopes), making it difficult to compare or draw conclusive results about their relative performance. With this benchmark, we aim to support researchers involved in developing new methods for localization and detection of biomolecular structures in cryo-electron tomograms.

Our contributions are:
\begin{itemize}
	\item We publish a new, publicly available, fully-annotated simulated cryo-electron tomography dataset. The dataset includes 12 protein classes, vesicles and gold fiducials.
	\item We evaluate and compare six learning-based methods and two versions of template matching.
	\item We note advantage of learning-based methods over template matching and show significant correlation between performance and molecular weight.
\end{itemize}

The remainder of this paper is organized as follows. Section 2 overviews dataset generation and benchmark evaluation. Then, in Section 3 we describe methods submitted for the evaluation. In Section 4 we present results. Finally, we discuss the results in Section 5.

\section{Benchmark}
\label{sec:benchmark}

We propose a task of localization and classification of particles in the cryo-electron tomogram volume. A benchmark is conducted on a simulated cryo-electron tomogram populated with randomly positioned and oriented copies of structurally well-defined molecular complexes. In total, the volume contained $1,571$ particles of $13$ different classes. To facilitate application of learning-based methods, we also provide nine tomograms with similar protein distribution and ground truth data that was used for the simulation.

\begin{table*}[]
\centering
\begin{tabular}{|l|l|l|l|l|l|l|}
\hline
\textbf{PDB} & \textbf{Name}       & \textbf{Mol. weight ($kDa$)} & \textbf{Volume ($nm^3$)} & \textbf{Area ($nm^2$)} & \textbf{Sphericity} & \textbf{Eff. radius ($nm$)} \\ \hline
1s3x         & Hsp70 ATPase        & 42.75                      & 90.82                                   & 109.8                                 & 0.89       & 2.481                     \\ \hline
3qm1         & LJ0536 S106A        & 62.62                      & 127.9                                   & 137.6                                 & 0.892      & 2.789                     \\ \hline
3gl1         & Ssb1, Hsp70         & 84.61                      & 196.5                                   & 191.2                                 & 0.855      & 3.083                     \\ \hline
3h84         & GET3                & 158.08                     & 347                                     & 370.9                                 & 0.644      & 2.807                     \\ \hline
2cg9         & Hsp90-Sba1          & 188.73                     & 401.2                                   & 358.4                                 & 0.734      & 3.358                     \\ \hline
3d2f         & Sse1p, Hsp70        & 236.11                     & 516                                     & 459.6                                 & 0.677      & 3.368                     \\ \hline
1u6g         & Cand1-Cul1-Roc1     & 238.82                     & 499.3                                   & 450.2                                 & 0.676      & 3.327                     \\ \hline
3cf3         & P97/vcp             & 541.74                     & 1136                                    & 745.2                                 & 0.707      & 4.573                     \\ \hline
1bxn         & Rubisco             & 559.96                     & 1021                                    & 583.4                                 & 0.84       & 5.25                      \\ \hline
1qvr         & ClpB                & 593.36                     & 1354                                    & 1063                                  & 0.557      & 3.821                     \\ \hline
4cr2         & 26S proteasome      & 1309.28                    & 2675                                    & 1846                                  & 0.505      & 4.347                     \\ \hline
5mrc         & Yeast mito ribosome & 3325.59                    & 6372                                    & 3161                                  & 0.526      & 6.047                     \\ \hline
\end{tabular}
\caption{Macromolecular complexes present in the dataset, sorted by their molecular weight. Volume, area, sphericity and effective radius are computed from particle volumes with threshold density $>0.5$.}
\label{tab:dataset_particles}
\end{table*}

\subsection{Dataset}
\label{sec:dataset}

First, we select $12$ proteins of known structure of varying size, shape and functions (Table~\ref{tab:dataset_particles}). To characterize them, we calculate sphericity, $\Psi$, a measure of how much the volume resembles a sphere:

\begin{equation}
\label{eq:sphericity}
\Psi = \frac{\pi^{1/3} \times (6V)^{2/3}}{A}
\end{equation}

and effective radius, the radius of a sphere with the same surface area to volume ratio as the volume of interest:

\begin{equation}
\label{eq:eff_radius}
r_eff = \frac{3V}{A}
\end{equation}

where $V$ is the volume and $A$ is the surface area.

For all molecules placed in the simulation, we first calculated an interaction potential. We define this potential as a sum with a real and imaginary part $V_{int} = V_{el} + i V_{ab}$. The electrostatic potential $V_{el}$ determines the elastic scattering of each molecule which influences phase contrast. We calculated it by placing on each atom's center a sum of 5 Gaussians that are parameterized by atom specific scattering factors \cite{rullgaard2011simulation}. We extended this electrostatic potential calculation scheme by correcting each atom for solvent exclusion \cite{fraser1978improved}. This was modelled by subtracting a smooth spherical volume around each atom with a Van der Waals radius determined by atom type, and an amorphous ice background potential of $4.530 V$. The second part of the potential - the absorption potential $V_{ab}$ - is dependent on molecule-type (i.e. protein, membrane, gold, or amorphous ice), and gives rise to absorption contrast through inelastic scattering \cite{vulovic2013image}.

We generated interaction potential maps of proteins, vesicles and gold fiducials at $5Å$. Then, without overlaps, we place $1,000$ to $3,000$ proteins, 7 to 14 gold fiducials and 2 to 7 vesicles at random locations and in random $SO(3)$ orientations, into the ground truth ``grandmodel'' - the box containing ground truth of the sample simulation. For each placed particle we save class, center coordinates and Euler angles in $ZXZ$ notation. That allows us to generate class masks (voxel to class mapping) and occupancy masks (voxel to particle mapping).

To simulate the embedding ice layer of the grandmodel, we added background constants of amorphous ice to the interaction potential of the grandmodel (corresponding to $4.530 V$ for the elastic part and 0.208 inelastic scattering fraction). These constants were both calculated assuming an amorphous ice density of $0.93 g/cm^3$, and for the absorption constant a $300kV$ electron beam. Each grandmodel was rotated over $61$ evenly spaced tilt angles ranging from $-60^{\circ}$ to $+60^{\circ}$, with cubic b-spline interpolation \cite{ruijters2012gpu} to minimize rotation artifacts.

To calculate the projection image for each rotation angle we implemented the multislice method \cite{vulovic2013image}. This method models the defocus gradient through the ice layer by propagating the electron wave through slices of the model. We set the size of these slices to $5 nm$. After calculating the wave propagation through the sample we obtain the exit wave in the image plane. To get the final projection image we multiplied the exit wave by the microscope’s contrast transfer function (CTF) and envelope functions using a defocus of $3.5 \mu m$ on average (see below), an acceleration voltage of $300 kV$, spherical aberration of $2.7 mm$, a source energy spread of $0.7 eV$, an illumination aperture of $30 /mu m$, and objective diameter of $100 /mu m$, a focal distance of $4.7 mm$, and no astigmatism. For the detection process we then convoluted the exit wave with the DQE of the K2SUMMIT detector. For the final electron counts, we sampled from the Poisson distribution, with an electron dose of approximately $1.6$  $e^{-} / \angstrom ^{2}$. The final images were $1024x1024$ pixels with a pixel size of $5Å$. We did a weighted back-projection reconstruction while binning the projections 2 times to obtain the final tomograms of $512x512x512$ with a sampling of $1 nm/voxel$. This means the initial models are oversampled compared to the final reconstruction, improving accuracy for the sample-microscope interaction. 

To introduce variation between tomograms we randomly selected a defocus between $2$ and $5 \mu m$ for each model, and an electron dose between $100$ and $120$  $e^{-} / \angstrom ^{2}$ for the full tilt range (which was equally divided over the 60 tilt images). We set random shifts for each projections in a $1 nm$ range in the $x$ and $y$ direction, to introduce misalignment of the projections and deteriorate reconstruction quality. This resulted in the tomograms varying in final SNR from 0.12 to 0.58 (the evaluation model 0.24), as calculated with $SNR = \sigma_{signal}^{2} / \sigma_{noise}^{2}$, where $\sigma_{signal}^{2} = \sigma_{noisy signal}^{2} - \sigma_{noise}^{2}$ (where the considered signal comes from all classes, including gold markers).

We noticed that in the power spectra (representing the amplitudes of the Fourier transform) of simulated projections, Thon rings were difficult to see. These are usually pronounced in experimental images (compared with data recorded at similar conditions; mixedCTEM from EMPIAR-10064). By scaling the amplitudes in Fourier space with amplitudes from experimental images (mixedCTEM) we qualitatively improved the appearance and made the simulation more similar to experimental images. We implemented this in a similar fashion to Fourier ring correlation (FRC), where for each ring of the Fourier amplitudes of the simulated image, we scaled the values to the mean of that same ring from an experimental image: $A_{scaled} = \sum_{i}^{N} M_{i} A_{sim} ( \mu_i^{exp} / \mu_i^{sim} )$, where $N$ is the number of rings, $M$ is the bandpass mask, and $\mu$ the mean of a band. We then obtained the updated simulated image by recombining the scaled amplitudes with the phase information of that image in Fourier space.

\subsection{Evaluation}
\label{sec:evaluation}

The main goal of the benchmark is to localize and classify biological particles in the tomographic reconstructions. The performance of the submissions has been evaluated solely on the test tomogram, the only tomogram for which ground truth is not available.

During evaluation, we parsed the submitted result and computed some commonly adopted performance metrics for classification and localization. The metrics are precision (Equation~\ref{eq:precision}): percentage of results which are relevant; recall (Equation~\ref{eq:recall}): percentage of total relevant results correctly classified; F1 score (Equation~\ref{eq:f1}): harmonic average of the precision and recall; false negative rate also known as miss rate (Equation~\ref{eq:fnr}): percentage of results which yield negative test outcomes. We also record how far the predicted center was from the ground truth center and how many results refer to the same particles.

\begin{equation}
\label{eq:precision}
\mathrm{Precision}=\frac{\mathrm{true~positive}}{\mathrm{true~positive} + \mathrm{false~positive}} \
\end{equation}

\begin{equation}
\label{eq:recall}
\mathrm{Recall}=\frac{\mathrm{true~positive}}{\mathrm{true~positive} + \mathrm{false~negative}} \
\end{equation}

\begin{equation}
\label{eq:f1}
\mathrm{F_1~score} = 2 \cdot \frac{\mathrm{precision} \cdot \mathrm{recall}}{\mathrm{precision} + \mathrm{recall}}
\end{equation}

\begin{equation}
\label{eq:fnr}
\mathrm{Miss~rate}= 1 - \mathrm{recall}
\end{equation}

\subsubsection{Erratum}
During evaluation stage of the benchmark we have discovered an error in the dataset. One of the classes, protein \texttt{4v94}, has been generated incorrectly and always appeared twice next to each other. Moreover, the center of such doubled-particle was in the empty space between them. The reason is that the PDB upload was a mirrored structure, while naturally the protein occurs as a single particle. While that does not present a problem for semantic segmentation approaches that are trained on class masks, it is a problem for approaches that use the center locations that we provide. For the fairness of benchmark we have decided to remove \texttt{4v94} protein from the evaluation completely. After the competition has finished, we have fixed reported center locations in the updated version of the dataset.

\subsection{Comparison with earlier benchmarks}
\label{sec:comparison_earlier_benchmarks}

Localization and classification of particles in cryo-ET is an open problem with major challenges due to the nature of imaging process and biological sample size. Previous editions of this benchmark \cite{gubins2019classification, gubins2020shrec} already provide some insight into automated localization and classification methods for cryo-ET. In this edition, we significantly improved dataset generation process (Section~\ref{sec:dataset}: multislice method, variated defocus and electron dose, Fourier scaling to experimental images) and introduced membranes as an additional semantic class.

\section{Participants and methods}
\label{sec:methods}
Nine international research groups registered to the track, of which seven submitted their results. Each participant could submit as many result sets as long as they present an interesting difference, e.g. different selection of hyperparameters for the same method. In total, the benchmark compares eight result sets obtained with seven different methods listed in this section.

\subsection{URFinder: Macromolecules localization using combined 3D UNet3+ and ResNet}
\label{method1}
\textbf{By:} \phantom{Xiao Wang, Daisuke Kihara}\\

\begin{wrapfigure}{l}{0.4\textwidth}
    \centering
    \includegraphics[width=.4\textwidth]{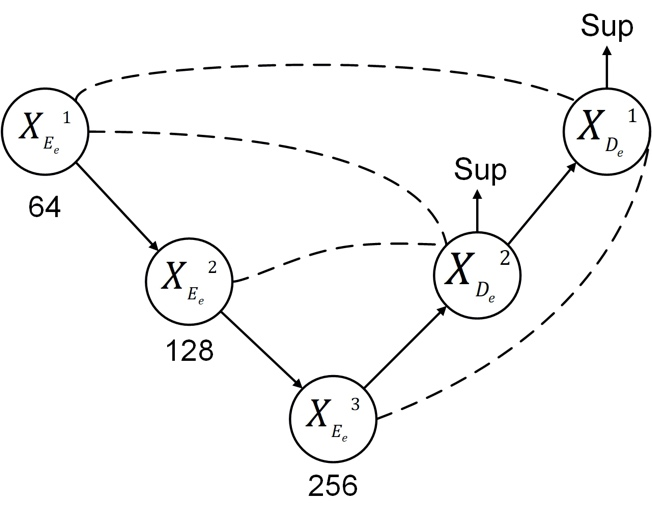}
    \caption{The architecture of 3D UNet3+.}
    \label{fig:urfinder}
\end{wrapfigure}

The method named URFinder is based on deep learning techniques, 3D UNet3+ \cite{huang2020unet} and 3D ResNet \cite{he2016deep, hara2017learning} for 3D semantic segmentation of the tomogram data. 3D UNet3+ was used to detect $13$ protein types. 3D ResNet was used for two classifications, one for detecting $13$ protein types and the other for detecting gold fiducials. For protein detection, results of 3D UNet3+ and 3D ResNet were combined.

Given a voxel (a cropped 3D region) from a tomogram, the proposed 3D-UNet3+ takes the voxel as input and outputs the $14$ probability scores, for $13$ proteins and background, for each grid point in the voxel. The size of each 3D input slice was set to $64x64x64$ and the stride size was set to $16$ to scan the whole cryo-ET map. We extended the original UNet3+, which was developed for 2D image, to 3D. In our architecture we have $2$ down-sampling and $2$ up-sampling operations instead of $4$ in the original UNet3+. The architecture of our 3D UNet3+ is shown in Figure~\ref{fig:urfinder}. All other configurations, such as convolution and maxpooling filter sizes, the number of filters and the stride size, are the same as UNet3+.

Two networks with 3D ResNet were trained. One is for protein detection, which is $14$ class ($13$ proteins and background) multi-class classification, and the other for binary classification for gold fiducials. 3D ResNet had $20$ layers \cite{huang2020unet, he2016deep} and the size of each 3D input slice was set to $32x32x32$. 
 	
We used tomograms 0 to 7 for training and validation while the tomogram 8 was kept for testing. For UNet3+ training, we scanned the whole map with a stride of $16$ with a voxel of $64x64x64$. We adopted the deep supervision technique in \cite{zhou2018unet++}. For output of each decoder, the binary cross-entropy (BCE) loss was applied and the total loss was defined as the sum of the individual losses. We used the Adam optimizer \cite{kingma2014adam} with an initial learning rate $0.0001$ and a weight decay of $1e^{-4}$. The cosine learning rate scheduler \cite{loshchilov2016sgdr} was used to decay the learning rate to $1e^{-5}$. For 3D ResNet training, we first sampled negative (background) examples from the provided training tomograms by extracting voxels with the center that is not closer than $16$ grid units to any proteins and gold fiducials. For each positive voxel that have a protein or a gold fiducial at the center, we augmented them by random flips and rotations. We had in total $620,892$ voxels for multi classification for detecting proteins. We had in total $2160$ positive voxels and $43916$ negative voxels for binary classification. In the training process, we used the Adam optimizer with learning rate of $0.002$ and a weight decay of $1e^{-4}$. In total, we train the network for $30$ epochs.

In the inference stage, we applied three networks, 3D UNet3+, 3D ResNet for protein detection and another ResNet for gold fiducial detection. 3D UNet3+ scanned the whole cryo-ET map with a stride of $16$ to extract $64x64x64$ voxels and made predictions at each grid point in the voxels. If a grid point has multiple assignments from different voxels, we used the average probabilities as the final predictions. We removed connected components with less than $64$ voxels and used centers of the rest for the reported protein particles. For 3D ResNet, we used a voxel of $32x32x32$ and a stride of $2$ to scan the whole map and made predictions to the center of the voxel. When detecting proteins, we checked the connected components and filtered out those components with less than $64$ voxels. For binary gold fiducial detection, too large components with width, height or length larger than $40$ voxels and too small components with less than $27$ voxels were removed.  Finally, detected proteins by 3D UNet3+ and 3D ResNet were combined. If detected proteins by the two methods overlap or locate in direct neighboring voxels they were merged into one protein.  The center of connected components were reported as the location of detected proteins and gold fiducials.

All the experiments were performed on NVIDIA Quodro RTX 8000 GPU. For 3D UNet3+, the training time was $7$ days for $30$ epochs on $2$ RTX 8000 GPU. The production time including inference and filtering was $72$ minutes on $1$ RTX 8000 GPU. For 3D ResNet, the training time was $4$ days for the multi-class and $1.5$ days for binary classification of gold fiducials for $30$ epochs on $1$ RTX 8000 GPU, and the production time was $54$ minutes including the inference and filtering.

\subsection{DeepFinder: Deep learning improves macromolecules localization and identification in 3D cellular cryo-electron tomograms}
\label{method2}
\textbf{By:} \phantom{Emmanuel Moebel}\\

\begin{figure*}[!t]
\centering
\includegraphics[width=0.925\textwidth]{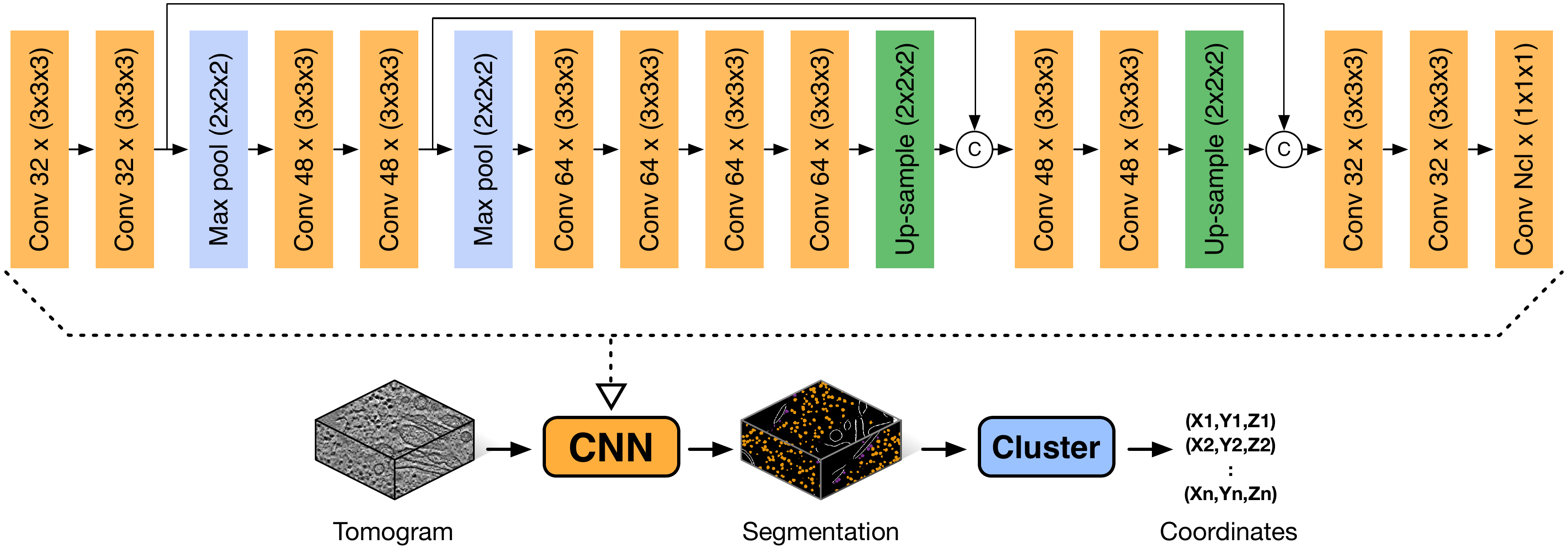}
\caption{Top: CNN architecture used in DeepFinder. All convolutional layers are followed by a ReLU activation function, except the last layer which uses a soft-max function. The up-sampling is achieved with up-convolutions (also called ``backward-convolution''). Combining feature maps from different scales is performed by concatenation along channel dimension.\\
Bottom: workflow depicting how macromolecule coordinates are obtained from the segmentations generated by the CNN. A clustering algorithm (mean-shift) is applied on the segmentation map to differentiate individual macromolecules.}
\label{fig:deep_finder}
\end{figure*}

\begin{figure}[h]
\centering
\includegraphics[width=.45\textwidth]{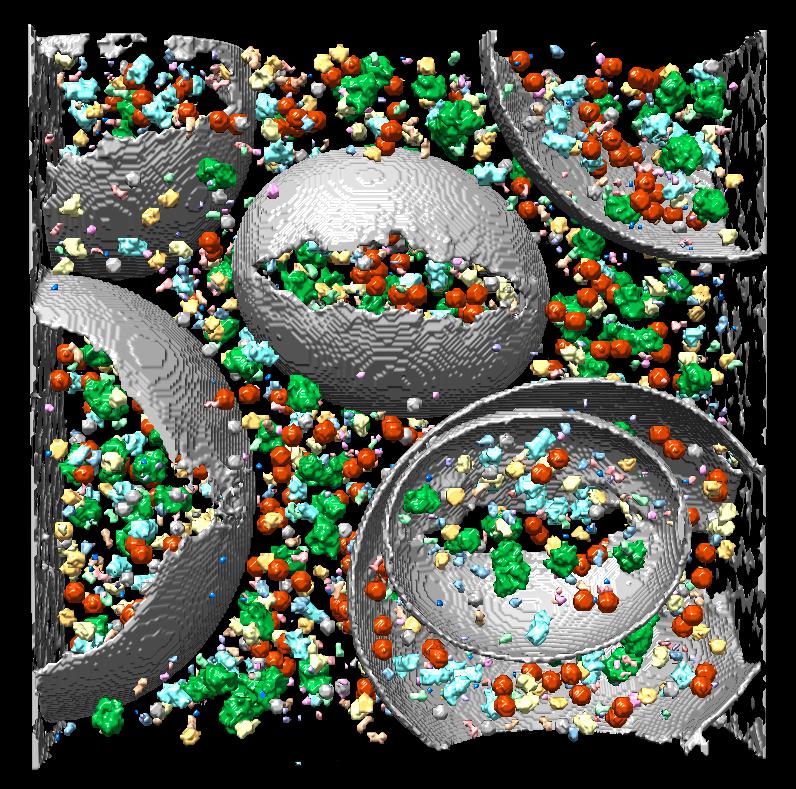}
\caption{Test tomogram segmentation with DeepFinder method.}
\label{fig:deep_finder_illustration}
\end{figure}

DeepFinder~\cite{moebel2020deep} is a computational tool for multiple macromolecular species localization, based on supervised deep learning. This two-step procedure (Figure~\ref{fig:deep_finder}) first produces a segmentation map where a class label is assigned to each voxel. The classes can represent different molecular species (e.g. ribosomes, ATPase), states of a molecular species (e.g. binding states, functional states) or cellular structures (e.g. membranes, microtubules). In the second step, the segmentation map is used to extract the positions of macromolecules. To perform image segmentation, we use a 3D CNN whose architecture and training procedure have been adapted for large datasets with unbalanced classes. The analysis of the obtained segmentation maps (Figure~\ref{fig:deep_finder_illustration}) is achieved by clustering the voxels with the same label class, using the mean-shift algorithm with different radii (bandwidth) for each class. Hence, the detected clusters correspond to individual macromolecules and their positions can then be derived. 

The 3D CNN architecture is trained with Adam~\cite{kingma2014adam} optimizer, using $0.0001$ as learning rate, $0.9$ as exponential decay rate for the first moment estimate and $0.999$ for the second moment estimate. A Dice loss~\cite{sudre2017generalised} is used to estimate the network parameters. The training took $50$ hours on an Nvidia M40 GPU. For large and medium macromolecules, presented scores are reached after $22$ hours; the additional time is necessary for having better performance with small macromolecules. The segmentation and clustering of a $512x512x200$ tomogram takes $20$ minutes.

With feasibility in mind, we developed training strategies to assist the user in producing segmentation maps (needed for training the CNN) from tomogram annotations consisting of the spatial coordinates of macromolecules. DeepFinder is an open-source python package with a graphical interface aimed towards non-computer scientist users.

\subsection{U-CLSTM: U-net architecture with convolutional long short term memory decoder}
\label{method3}
\textbf{By:} \phantom{Nguyen P. Nguyen, Tommi A. White, Filiz Bunyak}\\

\begin{figure*}[!t]
\centering
\includegraphics[width=0.925\textwidth]{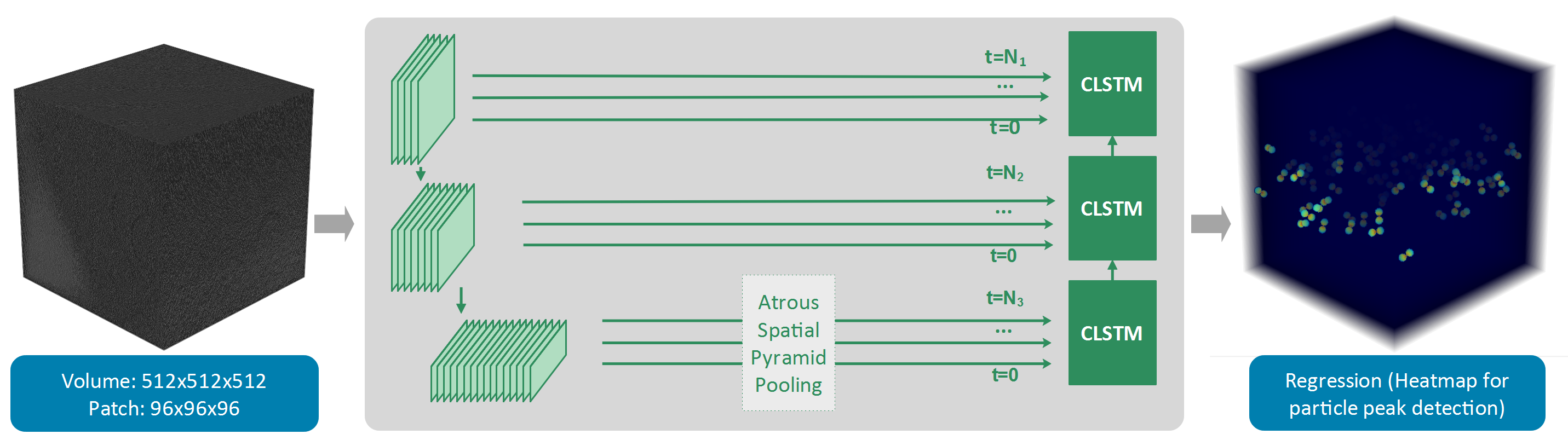}
\caption{U-CLSTM: U-net architecture with convolutional long short term memory (CLSTM) decoder}
\label{fig:uclstm}
\end{figure*}

To predict particle location, we employed the U-net architecture~\cite{ronneberger2015u} network. Our network's encoder has three main layers, each layer contains 20 residual blocks~\cite{zhang2017beyond}. It's not necessary for both the encoder and decoder to have the same configuration. We wanted to make use of the context memory mechanism~\cite{milletari2018cfcm} by using a convolutional long short term memory (CLSTM) cell~\cite{shi2015convolutional} in the decoder. This decoder architecture can exploit all image features from coarse to fine levels, further refine location prediction. The encoder and decoder are connected by the atrous spatial pyramid pooling block~\cite{chen2017deeplab}.
Instead of using mask segmentation to obtain particle centers, we applied regression to predict a heatmap of particle locations as in~\cite{nguyen2021drpnet}. Each heatmap contains $15$ channels, corresponding to $15$ types of particles to be detected. Different image volumes have different noise levels and different distributions of particles. We employed weighted sampling to balance the occurrence of data samples, and the small particles also have higher sampling weights. The ground truth heatmap was generated from a binary ground truth masks using distance transform. We then  used a mean squared error loss function to optimize the network parameters. U-CLSTM was trained with patches of size $96x96x96$ on an NVIDIA Quadro RTX-5000 GPU in $120$ hours. Total prediction time is $~15$ minutes for each image volume $512x512x512$, which includes both heatmap prediction time and particle center detection time.
Thresholding and connected component labeling were applied to each channel of the predicted heatmap to localize and segment the particle centers. Spurious detections were filtered out based on detection size. Particles whose centroids  are located within $5$ voxels from the  ground truth particle centroids in terms of Euclidean distance  are considered as detected. Detections who predict the same particle type as the corresponding  ground truth particle, are considered as correct classification.

\subsection{Multi-Cascade DS Network}
\label{method4}
\textbf{By:} \phantom{Giorgos Papoulias, Stavros Gerolymatos, Evangelia I. Zacharaki, Konstantinos Moustakas}\\

\begin{figure*}[!t]
\centering
\includegraphics[width=0.9\textwidth]{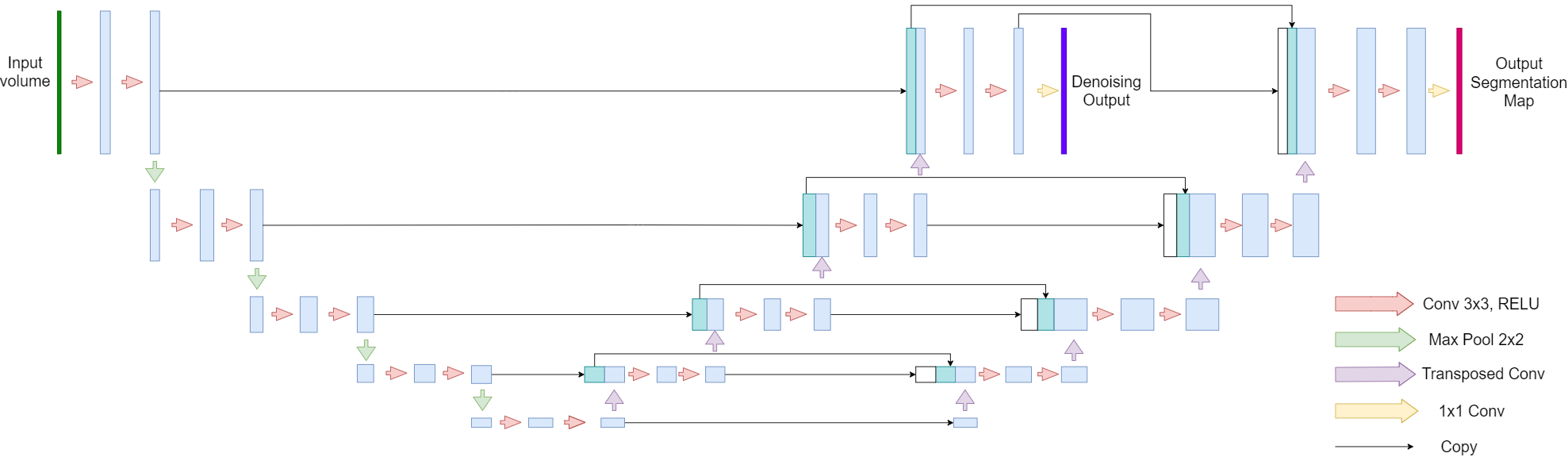}
\caption{Topology of Multi-Cascade DS Network}
\label{fig:multi_cascade_ds}
\end{figure*}

We formulated the classification and localization tasks as a supervised (volumetric) segmentation and morphological analysis problems, respectively. We solved the segmentation task jointly with denoising by employing a deep encoder-decoder architecture inspired by the cascaded network in~\cite{gubins2020deeply}. Specifically, we implemented a multi-cascade DS (Denoising-Segmentation) network based on the popular 3D U-Net~\cite{ronneberger2015u} and composed of two decoding pathways. The two pathways perform denoising on the input data
(3D tomogram) and volumetric segmentation (to produce a 3D label map), respectively. The whole architecture is illustrated in Figure~\ref{fig:multi_cascade_ds}.

In more details, in the denoising output pathway each decoder block is connected to the respective encoder layer with a skip connection, while in the segmentation output pathway
each block receives the skip connections from the respective layers of both the encoding pathway and the denoising decoding pathway. The connectivity introduced between decoding pathways is considered to facilitate inductive transfer between early and later stages of a deep cascade. Thus, this approach is more suitable than architectures dealing with denoising and segmentation in a serial fashion. Additionally, it is less computationally expensive as a serialized architecture would practically require the training of two deep networks, one for each learning task, independently.

We set the depth of the employed deep network to $5$ and the number of filters to $16$, $32$, $64$, $128$ and $256$ in the respective layers, yielding $13.57$ M of parameters in total. For the denoising task, we use the RMSE loss between the reconstruction and the respective grandmodel volumes and for 3D segmentation, we employ the Tversky loss function with $\alpha = 0.7$ and $\beta = 0.3$ using the ground truth segmentation masks as target. The unified loss function minimized during optimization includes the sum of the previous two loss terms. Loss minimization is performed using the Adam optimizer~\cite{kingma2014adam} using an initial $0.001$ learning rate explicitly defined by a Cosine Annealing learning rate scheduler. The model was trained for $20$ epochs using a batch size of $20$ on an NVIDIA GeForce RTX 3090 graphics card. PyTorch with CUDA acceleration was utilized for the implementation. Finally, the training procedure lasted $22$ hours and the inference time was approximately $5$ min.

Regarding the given dataset exploitation, the tomograms were cropped into cubic volumes (also denoted as subtomograms) with a size of $64^3$ using a $75\%$ overlap in all three dimensions. Half of the generated cubic volumes were horizontally and vertically flipped randomly during the training procedure for data augmentation purposes. Subtomograms from tomograms 0 to 7 were used for training and subtomograms from tomogram 8 for validation and optimization of hyper-parameters.

After having derived the 3D segmentation maps, connected component analysis (with neighborhood $26$) is performed to identify the individual particles as uniform clusters, followed by
two post-processing steps. First, spurious clusters that consist of less than $5$ voxels are removed and then classes are merged inside each component by assigning the most frequently occurring label to the whole component. Finally, the centroids of each component are estimated as the center of mass to address the localization challenge.

\subsection{YOPO: one-step object detection for cryo-ET macromolecule localization and classification}
\label{method5}
\textbf{By:} \phantom{Xiangrui Zeng, Sinuo Liu, Min Xu}\\

\begin{wrapfigure}{l}{0.45\textwidth}
    \centering
    \includegraphics[width=.45\textwidth]{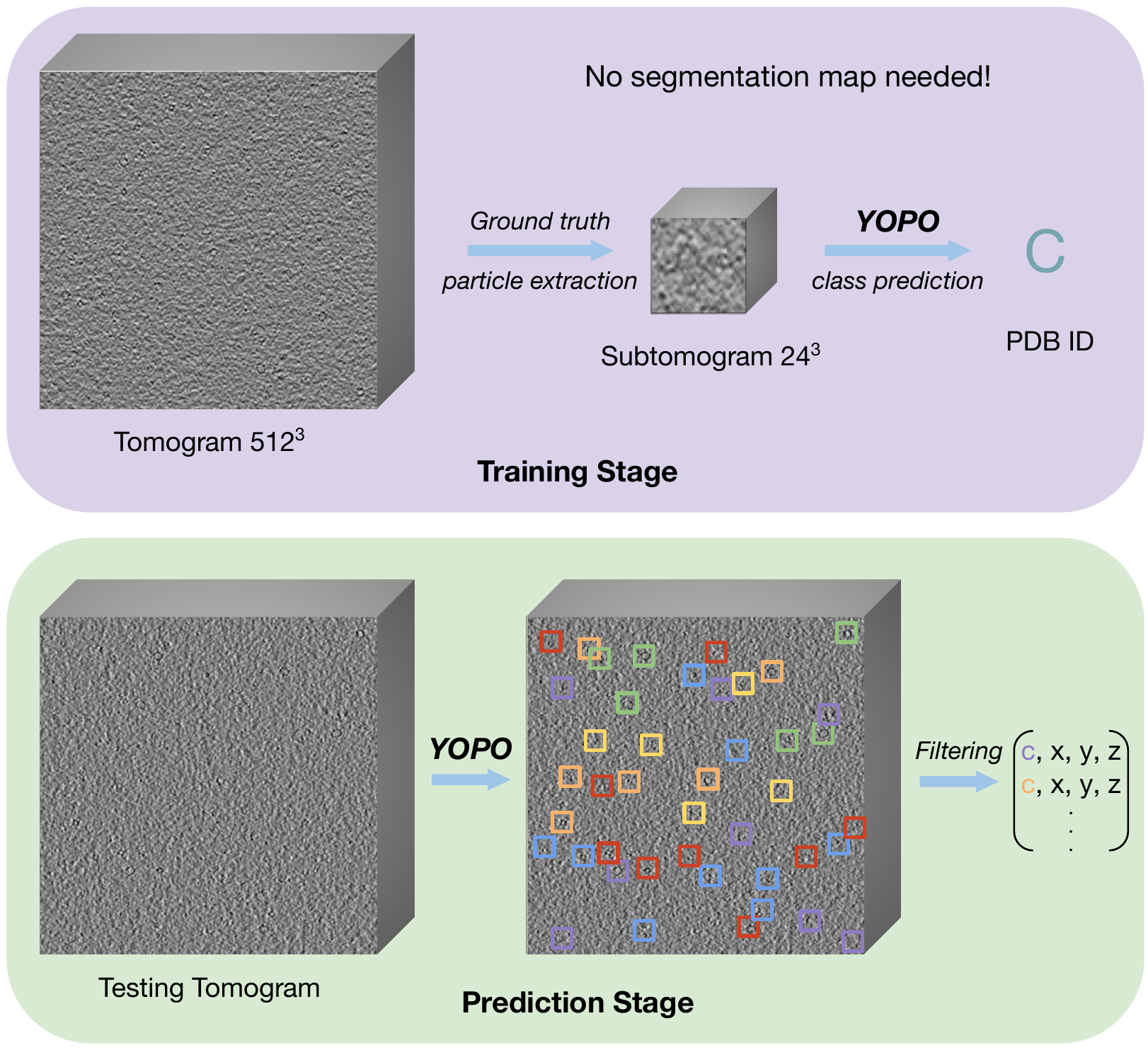}
    \caption{YOPO: Flowchart of macromolecule detection.}
    \label{fig:yopo}
\end{wrapfigure}

We formulate a novel one-step object detection framework specifically designed for cryo-ET data (Fig.~\ref{fig:yopo}). Previous deep learning-based works on detecting particles in cryo-electron tomograms are either two-step classification (extract potential structural regions as subtomograms and then perform classification) or segmentation methods. Considering two important properties of subtomogram data: (1) the high-level structural details of a particle determine its function and identity and (2) the particle is of random orientation and displacement inside a subtomogram, we designed a convolutional neural network named YOPO~\cite{zeng2021disca}, which retains discriminative high-level structural details and achieve the maximal transformation-invariance. The flowchart of macromolecule localization and classification using YOPO is illustrated in Fig.~\ref{fig:yopo}.

In the training stage, only particle location ground truth was used to train the YOPO network to predict the PDB ID of a subtomogram. In the testing stage, the trained YOPO network was applied on the tomogram level to directly predict both the location and PDB ID of detected macromolecules. From each training tomogram, we extract subtomograms of size $32^3$ according to the ground truth particle location file. An additional $20.000$ subtomograms were extracted at random locations from the background. Therefore, there are $K = 15$ classes in total including the background class and excluding vesicle centers. Subtomograms from tomogram $0$ to $7$ were used as training data and subtomograms from tomogram $8$ as validation data. The training took $8$ hours on one NVIDIA GeForce Titan X GPU. The trained model predicted at every location by applying the learned model parameters on the whole testing tomogram. Locations with high confidence (probability $> 0.9$) to be one of the structural classes were kept. We then filtered the locations to ensure that the minimum distance between two detections was greater than $14$ voxels. As a one-step object detection method, the classification and localization tasks are unified in an end-to-end fashion.

YOPO is an efficient cryo-ET macromolecule detection (localization + detection) framework in two aspects: (1) the only ground truth information used for training is the particle locations and classes in ground truth particle location file; (2) YOPO performs prediction on the subtomogram level at every location, which is similar to the traditional template matching approach. However, the whole prediction on one tomogram took only about $40$ min using one GPU instance.

\subsection{Central Feature Network (CFN) for cryo-EM particle classification and localization}
\label{method6}
\textbf{By:} \phantom{Yaoyu Wang, Cheng Chen, Fa Zhang, Xuefeng Cui}\\

\begin{figure}[!t]
\centering
\includegraphics[width=.45\textwidth]{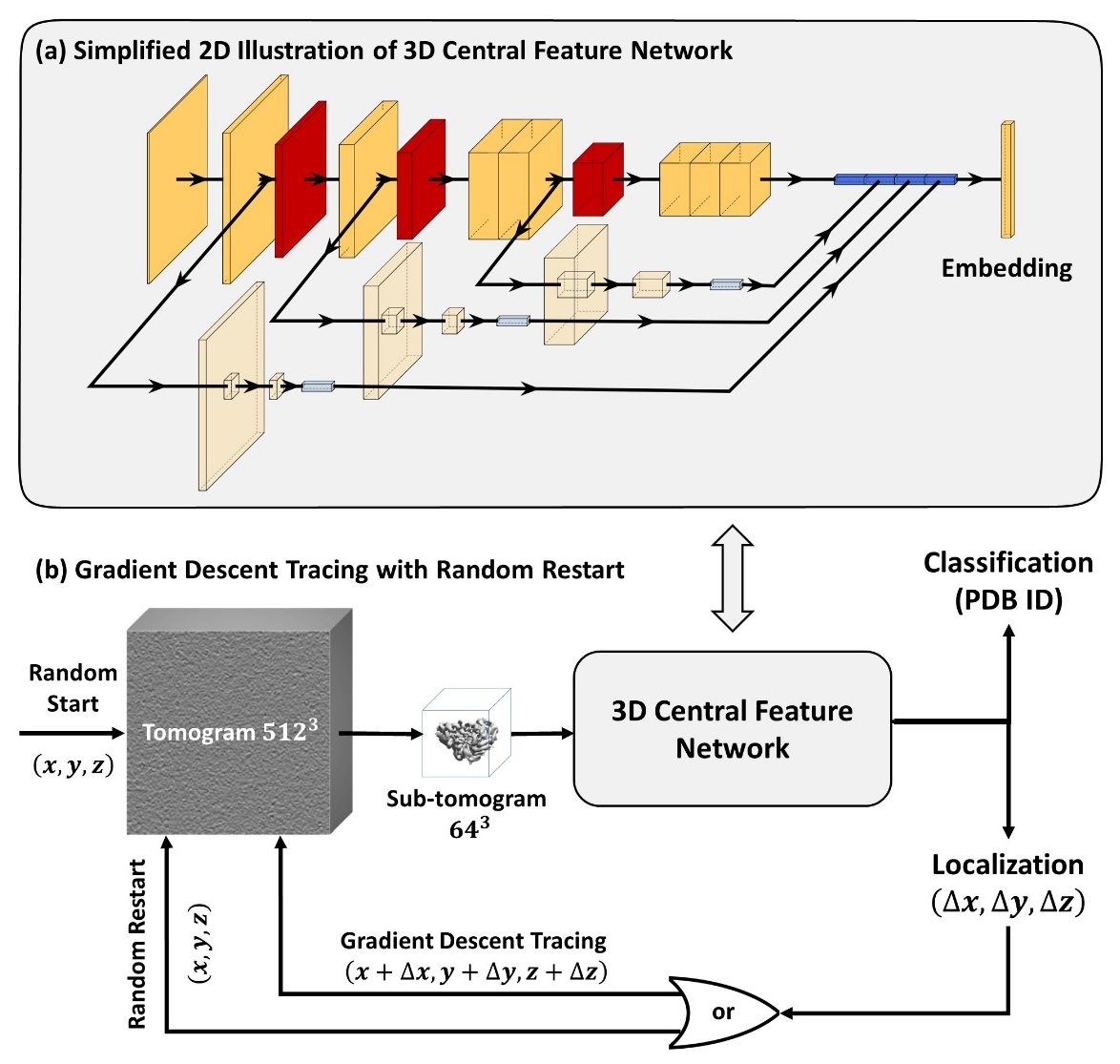}
\caption{Central Feature Network (CFN) architecture and inference pipeline}
\label{fig:cfn}
\end{figure}

We introduced a novel Central Feature Network (CFN) for the general 3D object detection problem, and applied it on the Cryo-EM particle detection problem. Specifically, CFN takes $64x64x64$ sub-tomograms as input, and detects particles in the input sub-tomograms. As shown in Figure 1a, CFN is based on a 3D ResNet model~\cite{he2016deep, he2016identity} with dilated convolutions~\cite{yu2015multi} and the focal loss function~\cite{lin2017focal}. Notably, our CFN model is different from existing models with three novel modifications. First, existing methods use only the neurons of the last convolutional block for predictions, while we combine the central neurons (i.e., blue boxes in Figure~\ref{fig:cfn}a) of each convolutional block for predictions. This helps to identify particles with different sizes because deeper networks are more suitable for bigger particles while shallower networks are more suitable for smaller particles. Second, existing methods use average pooling (or max pooling) for dimension reduction, while we use bottlenecked pooling (i.e., red boxes in Figure~\ref{fig:cfn}a) with two fully connected layers for the same task. By avoiding location irrelevant operations (e.g., average or max), location information could be retained from layers to layers. Finally, existing methods focus on only classifications, while we perform both classifications and localizations at the same time.

As shown in Figure~\ref{fig:cfn}b, the localization results of CFN can be used to trace the particle centers via a gradient decent approach with a random restart. Specifically, an initial sub-tomogram is randomly sampled, and CFN is used to detect the particle in the sub-tomogram. If a particle is detected, the predicted particle center is used to sample another sub-tomogram. This center tracing process is repeated with a random restart so that all sub-tomograms can be sampled theoretically, while the sub-tomograms near particle centers are more likely to be sampled. Finally, near-center predictions are clustered to produce a consensus prediction. The training and testing process took four days on two NVIDIA 3090 video cards.

\subsection{Template matching}
\label{method7}
\textbf{By:} \phantom{M. Cristina Trueba, Marten L. Chaillet}\\

We performed template matching on the simulated dataset using the cryo-ET analysis framework PyTom~\cite{hrabe2012pytom}. A solvent corrected electrostatic potential sampled to a grid of $1$ $nm$ voxels was modulated for each particle with a CTF curve at 3,65 $\mu m$ defocus in the frequency domain to serve as templates~\cite{fraser1978improved}. The templates were flipped to cover left- and right-handedness in the particle orientation of the simulated dataset. In addition, a Gaussian low pass filter set to 4 nm was applied in Fourier space to both the template and the tomogram to increase contrast and facilitate the particle detection. Spherical template masks with Gaussian smoothed edges were used for normalization of the cross-correlation value. The masks radius for each particle was chosen to fully encompass the template. We also used Laplace of Gaussian (LoG) with standard deviation ($\sigma$) of $5$ to recognize gold markers and extract their position. Template matching for $12$ protein classes and both handedness takes 4 hours and 26 minutes on the NVIDIA GeForce GTX 1080 Ti (20 min 40 s per class, 10 min 20 s per handedness).

\textbf{TM.} The top $1,000$ candidates with the highest cross-correlation score for each class were extracted and merged for the different particle handedness. A Gaussian distribution was fitted to the histogram of the correlation scores for each case pursuing to identify the correct particle population in it. We objectively set a minimum threshold to the mean ($\mu$) minus two times the standard deviation ($\sigma$) of the fitted gaussian population to avoid false positives. This resulted in $9$ fiducials and $122, 107, 318, 314, 924, 708, 211, 329, 127, 399, 149, 301$ particles of each class in the gold marker excluded dataset, from largest to smallest with a total of $4,009$ particles recognized.

\textbf{TM-F.} Alternatively, candidates of each class were additionally filtered to exclude those that would potentially be overlapped with already selected particles. To test for overlap, we calculate the distance between the centre of an existing particle to the centre of the candidate and calculate whether the distance is smaller than the sum of their radii. The candidates for round and symmetrical particles were filtered before asymmetrical and elongated particles as their performance was significantly better based on visual inspection in PyTom.  This resulted in 9 fiducials and $122, 81, 79, 127, 624, 212, 37, 71, 125, 85, 32, 65$ particles of each class recognized in the gold marker excluded dataset, from largest to smallest respectively with a total of $1,660$ particles selected.

\section{Results}
\label{sec:results}

We have evaluated different metrics (Section~\ref{sec:evaluation}) that allows comparison of localization (Table~\ref{tab:local_eval}) and classification (Table~\ref{tab:class_eval}) performance of the methods. For more convenient referencing, we have assigned following short names to the methods:

\begin{enumerate}
	\item URFinder (Section~\ref{method1})
	\item DeepFinder (Section~\ref{method2})
	\item U-CLSTM (Section~\ref{method3})
	\item MC DS Net (Section~\ref{method4})
	\item YOPO (Section~\ref{method5})
	\item CFN (Section~\ref{method6})
	\item TM-T and TM-F (Section~\ref{method7})
\end{enumerate}

\begin{figure}[!t]
\centering
\includegraphics[width=.48\textwidth]{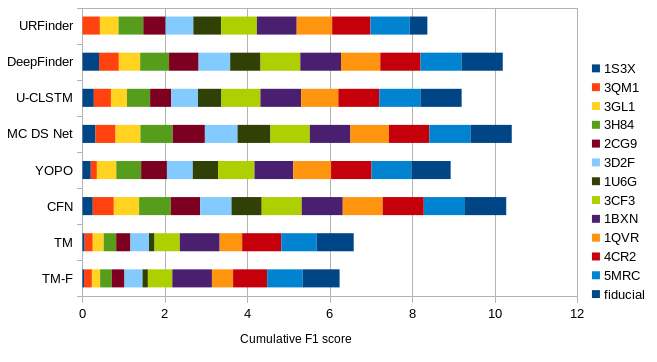}
\caption{Cumulative classification F1 scores of methods.}
\label{fig:cumulative_class_f1}
\end{figure}

The test tomogram has $1,571$ particles of the same $13$ classes and same distribution as the training data (Table~\ref{tab:particles_test_distribution}). To have a more detailed classification evaluation, we compare results with cumulative F1 score (Figure~\ref{fig:cumulative_class_f1}), as well as group proteins by their molecular weight (Table~\ref{tab:group_by_weight}) and average F1 scores for an additional metric correlated with particle sizes (Table~\ref{tab:f1_by_weight}).

\begin{table*}[h]
\centering
\begin{tabular}{|c|c|}
\hline
\textbf{Particle} & \textbf{Quantity} \\ \hline
\texttt{1s3x}         & 122               \\ \hline
\texttt{3qm1}         & 120               \\ \hline
\texttt{3gl1}         & 123               \\ \hline
\texttt{3h84}         & 144               \\ \hline
\texttt{2cg9}         & 125               \\ \hline
\texttt{3d2f}         & 140               \\ \hline
\texttt{1u6g}         & 143               \\ \hline
\texttt{3cf3}         & 139               \\ \hline
\texttt{1bxn}         & 135               \\ \hline
\texttt{1qvr}         & 127               \\ \hline
\texttt{4cr2}         & 115               \\ \hline
\texttt{5mrc}         & 121               \\ \hline
fiducial              & 11                \\ \hline
\end{tabular}
\caption{Distribution of particles by class in the test tomogram.}
\label{tab:particles_test_distribution}
\end{table*}

\begin{table*}[]
\centering
\begin{tabular}{|l|l|l|l|l|l|l|l|l|l|l|}
\hline
\textbf{Method}     & \textbf{RR} & \textbf{TP} & \textbf{FP} & \textbf{FN} & \textbf{MH} & \textbf{AD} & \textbf{Recall} & \textbf{Precision} & \textbf{Miss rate} & \textbf{F1} \\ \hline
URFinder   & 1969        & 1298        & 377         & 267         & 149         & 1.84        & 0.826           & 0.659              & 0.174              & 0.733       \\ \hline
DeepFinder & 1567        & 1362        & 64          & 203         & 20          & 2.22        & 0.867           & 0.\textbf{869}              & 0.133              & \textbf{0.868}       \\ \hline
U-CLSTM    & 1460        & 1253        & \textbf{49}          & 312         & 44          & 2.13        & 0.798           & 0.858              & 0.202              & 0.827       \\ \hline
MC DS Net  & 1760        & \textbf{1415}        & 239         & \textbf{150}         & 56          & 1.59        & \textbf{0.901}           & 0.804              & \textbf{0.099}              & 0.850       \\ \hline
YOPO       & 1627        & 1224        & 232         & 341         & \textbf{14}          & 1.66        & 0.720           & 0.752              & 0.221              & 0.765       \\ \hline
CFN        & 1765        & 1364        & 239         & 201         & 20          & \textbf{1.52}        & 0.868           & 0.773              & 0.132              & 0.818       \\ \hline
TM-F       & 1772        & 963         & 295         & 601         & 17          & 2.65        & 0.613           & 0.543              & 0.387              & 0.576       \\ \hline
TM         & 4195        & 1073        & 583         & 492         & 716         & 2.62        & 0.683           & 0.256              & 0.317              & 0.372       \\ \hline
\end{tabular}
\caption{Results of localization evaluation. \textit{RR}: results reported; \textit{TP}: true positive, unique particles found; \textit{FP}: false positive, reported non-existant particles; \textit{FN}: false negative, unique particles not found; \textit{MH}: multiple hits: unique particles that had more than one result; \textit{AD}: average euclidean distance from predicted particle center in voxels; \textit{Recall}: uniquely selected true locations divided by actual number of particles in the test tomogram; \textit{Precision}: uniquely selected true locations divided by RR; \textit{Miss rate}: percentage of results which yield negative results; \textit{F1 Score}: harmonic average of the precision and recall. The best results in each column are highlighted.}
\label{tab:local_eval}
\end{table*}

\begin{table*}[]
\centering
\begin{tabular}{|l|l|l|l|l|l|l|l|l|l|l|l|l|l|}
\hline
\textbf{Method} & \textbf{{1s3x}} & \textbf{{3qm1}} & \textbf{{3gl1}} & \textbf{{3h84}} & \textbf{{2cg9}} & \textbf{{3d2f}} & \textbf{{1u6g}} & \textbf{{3cf3}} & \textbf{{1bxn}} & \textbf{{1qvr}} & \textbf{{4cr2}} & \textbf{{5mrc}} & \textbf{{fiducial}} \\ \hline
URFinder        & 0.000         & 0.423         & 0.453         & 0.600         & 0.542         & 0.672         & 0.673         & 0.867         & 0.967         & 0.860         & 0.926         & 0.954         & 0.429             \\ \hline
DeepFinder      & \textbf{0.402}         & 0.481         & 0.517         & 0.701         & 0.716         & 0.766         & 0.737         & 0.964         & 0.989         & 0.953         & 0.974         & 0.996         & \textbf{1.000}             \\ \hline
U-CLSTM         & 0.277         & 0.415         & 0.389         & 0.561         & 0.511         & 0.651         & 0.566         & 0.946         & 0.989         & 0.903         & 0.991         & \textbf{1.000}         & \textbf{1.000}             \\ \hline
MC DS Net       & 0.316         & 0.487         & 0.603         & \textbf{0.783}         & \textbf{0.782}         & \textbf{0.791}         & \textbf{0.797}         & 0.956         & 0.985         & 0.934         & 0.979         & \textbf{1.000}         & \textbf{1.000}             \\ \hline
YOPO            & 0.203         & 0.148         & 0.471         & 0.601         & 0.626         & 0.627         & 0.613         & 0.884         & 0.938         & 0.920         & 0.983         & 0.966         & 0.952             \\ \hline
CFN             & 0.250         & \textbf{0.511}         & \textbf{0.613}         & 0.768         & 0.714         & 0.761         & 0.731         & \textbf{0.971}         & \textbf{0.996}         & \textbf{0.969}         & \textbf{0.996}         & \textbf{1.000}         & \textbf{1.000}             \\ \hline
TM-F            & 0.040         & 0.189         & 0.200         & 0.282         & 0.308         & 0.439         & 0.129         & 0.592         & 0.962         & 0.513         & 0.827         & 0.857         & 0.900             \\ \hline
TM              & 0.054         & 0.197         & 0.266         & 0.302         & 0.345         & 0.452         & 0.133         & 0.615         & 0.966         & 0.545         & 0.950         & 0.857         & 0.900             \\ \hline
\end{tabular}
\caption{Results of classification evaluation for all classes. The values correspond to F1 score achieved by methods on specific classes. The best results in each column are highlighted.}
\label{tab:class_eval}
\end{table*}

\begin{table*}[]
\centering
\begin{tabular}{|l|l|l|}
\hline
\textbf{Group} & \textbf{Weight} & \textbf{Proteins}            \\ \hline
Small          & \textless{}200  & 1s3x, 3qm1, 3gl1, 3h84, 2cg9 \\ \hline
Medium         & 200-600         & 3d2f, 1u6g, 3cf3, 1bxn, 1qvr \\ \hline
Large          & 600+            & 4cr2, 5mrc                   \\ \hline
\end{tabular}
\caption{Grouping of macromolecular complexes by their molecular weight in $kDa$}
\label{tab:group_by_weight}
\end{table*}

\begin{table*}[]
\centering
\begin{tabular}{|l|l|l|l|}
\hline
\textbf{Method} & \textbf{Small} & \textbf{Medium} & \textbf{Large} \\ \hline
URFinder        & 0.404          & 0.808           & 0.94           \\ \hline
DeepFinder      & 0.563          & 0.882           & 0.985          \\ \hline
U-CLSTM         & 0.431          & 0.811           & 0.996          \\ \hline
MC DS Net       & \textbf{0.594} & \textbf{0.893}  & 0.989          \\ \hline
YOPO            & 0.41           & 0.796           & 0.974          \\ \hline
CFN             & 0.571          & 0.886           & \textbf{0.998} \\ \hline
TM-F            & 0.204          & 0.527           & 0.842          \\ \hline
TM              & 0.233          & 0.542           & 0.903          \\ \hline
\end{tabular}
\caption{F1 scores of each submission for size group defined in Table~\ref{tab:group_by_weight}. The best results in each column are highlighted.}
\label{tab:f1_by_weight}
\end{table*}

\begin{figure*}[!t]
\centering
\includegraphics[width=\textwidth]{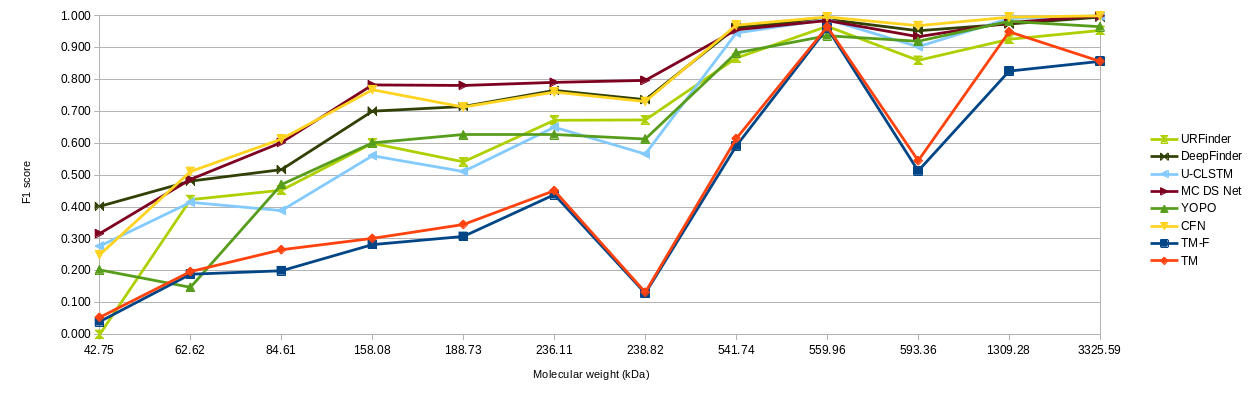}
\caption{Method classification performance plot against particle molecular weight.}
\label{fig:classification_vs_weight}
\end{figure*}

\begin{table*}[]
\centering
\begin{tabular}{|l|l|l|}
\hline
\textbf{Method} & \multicolumn{1}{l|}{\textbf{Training stage}} & \multicolumn{1}{l|}{\textbf{Inference stage}} \\ \hline
URFinder        & 300h                    & 2h 6m                    \\ \hline
DeepFinder      & 50h                     & 20m                      \\ \hline
U-CLSTM         & 120h                    & 15m                      \\ \hline
MC DS Net       & 22h                     & 5m                       \\ \hline
YOPO            & 8h                      & 40m                      \\ \hline
CFN             & \multicolumn{2}{c|}{96h}                                                                     \\ \hline
TM-F/TM GPU     & N/A                     & 4h 26m                   \\ \hline
\end{tabular}
\caption{Reported training and inference stages timings.}
\label{tab:method_timings}
\end{table*}

\begin{figure*}[]
\centering
\includegraphics[width=\textwidth]{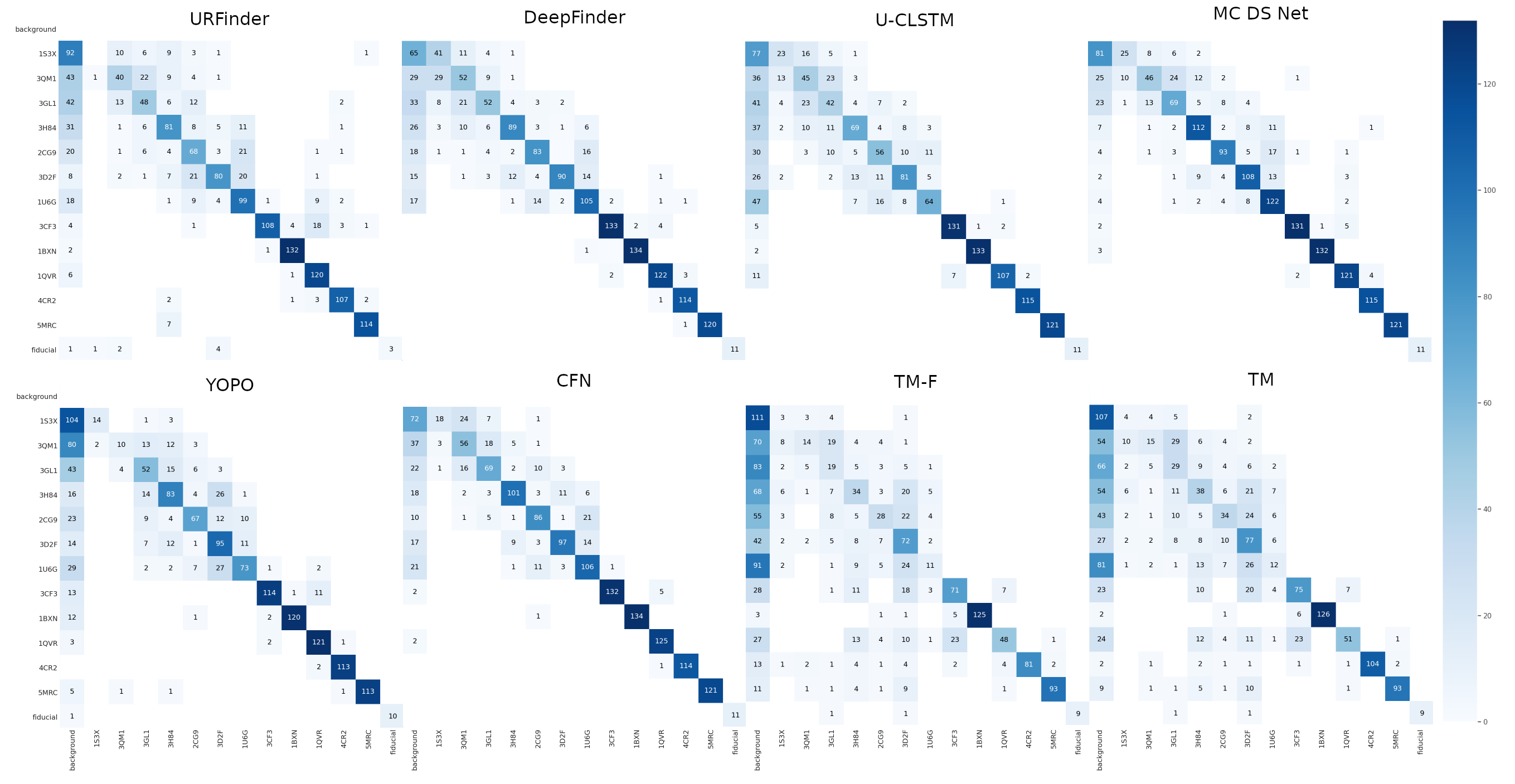}
\caption{Classification confusion matrices of the compared methods. The particles are ordered by molecular weight. The colorbar indicates the number of correct classifications.}
\label{fig:confusion_matrix}
\end{figure*}

\section{Discussion}
\label{sec:discussion}

The benchmark allowed us to compare baseline template matching and upcoming learning-based methods, as well as highlight current challenges and approaches in cryo-ET.

\subsection*{Learning-based vs. template matching}
The results (Table~\ref{tab:local_eval},~\ref{tab:class_eval}) show that all learning-based methods achieve better scores than the traditional baseline template matching (TM). Learning-based methods are also at least twice faster (Table~\ref{tab:method_timings}) than optimized GPU-accelerated TM, not taking into account ``offline'' training time. The success suggests that existing supervised models can do better than TM in practice, given the data is sufficiently realistic and/or model is robust to image acquisition parameter difference. Moreover, some unsupervised approaches~\cite{zeng2021disca, zeng2020gum} already show comparable or better performance on real datasets.

\subsection*{Localization precision}
Localizing exact center of a particle is important for accurate subtomogram averaging. During localization evaluation we have recorded average euclidean distance from predicted to ground truth particle center. CFN (Section~\ref{method6}) showed the best performance, closely followed by MC DS Net (Section~\ref{method4}) and YOPO (Section~\ref{method5}), again better than template matching almost on one full voxel (1nm). CFN and YOPO receive subtomograms as input and both use smart pooling approaches to maximize scale-invariance, allowing to accurately find bioparticles of different sizes. MC DS Net uses denoising that can remove noise around particles leading to improved localization precision.

\subsection*{Neural network architectures}
Four methods (DeepFinder, U-CLSTM, MC DS Net and partly URFinder) use advanced variations of U-Net~\cite{ronneberger2015u} architecture, originally intended for accurate, voxel-level, biomedical semantic segmentation. CFN and YOPO do not rely on semantic segmentation rather work with subtomograms directly and do not require voxel level labels, making it more accesible for cryo-ET researchers.

\subsection*{Performance correlates with molecular weight}
Results (Table~\ref{tab:f1_by_weight}, Figure~\ref{fig:classification_vs_weight},  Figure~\ref{fig:confusion_matrix}) show strong correlation between molecular weights and classification performance for all methods. All learning-based methods show consistent performance directly correlating with size, probably due to voxel count (volume) going down rapidly and not leaving enough voxels to be classified. At the same time, TM results have interesting difference, being able to distinguish some particles better than other. For example, TM performs on-par with learning-based method with protein \texttt{1bxn} (rubisco). Rubisco has high sphericity, large effective radius (Table~\ref{tab:dataset_particles}) and four-fold symmetry, that fits well to template matching process. On the opposite, protein \texttt{1u6g} is asymmetric, has average sphericity and effective radius, and is not distinguished well by TM.

\subsection*{Future work}
We strive to provide highly realistic dataset, and while the simulator shows good agreement with experimental data, it has not been quantitatively validated yet. We hope that in the next edition of the benchmark we can provide a fully annotated experimental tomogram as one of the test objectives.

\section*{Acknowledgments}
\phantom{We want to thank all track participants for their contribution.}\\
\phantom{This work was supported by the European Research Council under the European Union’s Horizon2020 Programme (ERC Consolidator Grant Agreement 724425 - BENDER) and the Nederlandse Organisatie voor Wetenschappelijke Onderzoek (Vici 724.016.001 and 741.018.201).}

\end{document}